\documentclass[preprint,sort&compress,number,3p,twocolumn]{elsarticle} 

\usepackage{lineno}
\usepackage{amsmath}
\usepackage{booktabs}
\usepackage{graphicx}

\begin{document}

\begin{frontmatter}

\title{A Bayesian technique for improving the sensitivity of the\\atmospheric neutrino L/E analysis}

\author[label1]{A. S. T. Blake\corref{cor1}}
\author[label1]{J. D. Chapman}
\author[label1]{M. A. Thomson}

\cortext[cor1]{Corresponding author}

\address[label1]{Cavendish Laboratory, JJ Thomson Avenue, Cambridge, CB3 0HE, United Kingdom}

\date{\today}

\begin{abstract}
This paper outlines a method for improving the precision
of atmospheric neutrino oscillation measurements.
One experimental signature for these oscillations is an 
observed deficit in the rate of $\nu_{\mu}$ charged-current
interactions with an oscillatory dependence on $L_{\nu}/E_{\nu}$,
where $L_{\nu}$ is the neutrino propagation distance,
and $E_{\nu}$ is the neutrino energy. 
For contained-vertex atmospheric neutrino interactions, 
the $L_{\nu}/E_{\nu}$ resolution varies significantly 
from event to event.
The precision of the oscillation measurement can be improved by 
incorporating information on $L_{\nu}/E_{\nu}$ resolution into 
the oscillation analysis. In the analysis presented here, 
a Bayesian technique is used to estimate the $L_{\nu}/E_{\nu}$ 
resolution of observed atmospheric neutrinos on an event-by-event basis.
By separating the events into bins of $L_{\nu}/E_{\nu}$ resolution in the oscillation analysis, 
a significant improvement in oscillation sensitivity can be achieved.
\end{abstract}

\begin{keyword}
Neutrino oscillations \sep atmospheric neutrinos
\end{keyword}

\end{frontmatter}


\section{Introduction}
\label{Introduction}

It has now been firmly established by experiment that atmospheric
neutrinos undergo oscillations between flavours.
In the standard theory, the oscillations arise from quantum
mechanical mixing between the neutrino flavour eigenstates.
This is governed by a unitary PMNS mixing matrix~\cite{neutrinomixing1,neutrinomixing2,neutrinomixing3}, 
which can be parameterised using three mixing angles and a CP-violating phase.
The amplitudes of the oscillations depend on the sizes of the mixing angles,
whereas the wavelengths depend on the neutrino squared-mass splittings,
$\Delta m^{2}_{ji}=m^{2}_{j}-m^{2}_{i}$,
and also on $L_{\nu}/E_{\nu}$, where $L_{\nu}$ is the neutrino 
propagation distance and $E_{\nu}$ is the neutrino energy. 

The observed atmospheric neutrino $\nu_{\mu}$ data are presently
well-described by an effective two-flavour model of vacuum oscillations.
In this approximation, the survival probability of an initial state 
$|\nu_{\mu}\rangle$ is given by:

 \begin{equation}
  P \left( \nu_{\mu} \rightarrow \nu_{\mu} \right) 
     = 1 - \sin^{2}2\theta 
         \sin^{2}\left( \frac{1.27 \Delta m^{2}(\mbox{eV}^{2}) 
           L_{\nu}\mbox{(km)} }{E_{\nu}\mbox{(GeV)} } \right).
 \label{OscFormula}
 \end{equation}

The two-flavour oscillation parameters,
$\Delta m^{2}$ and $\sin^{2}2\theta$,
have been measured in atmospheric neutrinos by a number of experiments, 
including Super-Kamiokande~\cite{superk1,superk2,superk3},
MACRO~\cite{macro}, Soudan~2~\cite{soudan2} 
and MINOS~\cite{minoscontainedvertex,minosupwardmuon,minosatmos}.
In addition, the atmospheric neutrino results have been 
confirmed by the K2K~\cite{k2k}, MINOS~\cite{minos1,minos2,minos3} and T2K~\cite{t2k}
long-baseline experiments, using accelerator beams of muon neutrinos. 
The MINOS experiment reports confidence limits of 
$|\Delta m^{2}| = (2.32^{+0.12}_{-0.08}) \times10^{-3} \mbox{eV}^{2}$ (68\% C.L.)
and $\sin^2 2\theta > 0.90$ (90\% C.L.)~\cite{minos3}.

For atmospheric neutrino experiments, one signature 
of neutrino oscillations is a deficit in the observed
rate of $\nu_{\mu}$ charged-current (CC) interactions
relative to the prediction without oscillations.
The size of the deficit varies with $L_{\nu}/E_{\nu}$
according to the formula given in Eq.~\ref{OscFormula}.
The $\nu_{\mu}$ CC interactions are identified by the presence 
of an emitted muon, which may also be accompanied 
by additional shower activity produced by the final-state hadronic system. 
The neutrino kinematics can be reconstructed by combining 
measurements of the emitted muon and hadronic system,
obtained by analysing the observed hits in the detector.
This yields measured values for the neutrino energy, $E_{\nu}$, 
and zenith angle, $\theta_{\nu}$. 
The reconstructed zenith angle 
is converted into a propagation distance, $L_{\nu}$,
using a model of atmospheric neutrino production height.

The precision with which an atmospheric neutrino experiment 
can measure the oscillation parameters is dependent
on its resolution of the neutrino energy and direction. 
Typically, the energy and direction of the muon 
can be measured precisely. However, the measurement 
of the hadronic system has a worse resolution.
Therefore, the resolution of both $E_{\nu}$ and $\theta_{\nu}$
improves with $E_{\mu}/E_{\nu}$, where $E_{\mu}$ is the muon energy. 
The angular resolution also improves with neutrino energy
since the final-state particles are increasingly aligned 
with the incident neutrino direction.
This is particularly important when the reconstructed 
muon direction is used to approximate the neutrino direction.
The overall resolution is dependent on the precise configuration 
of the detector geometry. For example, 
the resolution is worse if there are fewer hits in the detector, 
or if the final-state particles are not fully contained in the detector.

The resulting resolution of $L_{\nu}/E_{\nu}$ varies 
significantly across the $\nu_{\mu}$ CC data sample.
For the measurement of $E_{\nu}$, the resolution is
worse for neutrino interactions with higher inelasticity,
since the muon kinematics are better measured
than those of the hadronic system.
For the measurement of $L_{\nu}$, the resolution is
worse at lower neutrino energies, 
where the average scattering angle between the 
neutrino and muon is larger,
and also worse at the horizon, 
where $L_{\nu}$ varies rapidly as a function of zenith angle,
so that a small uncertainty in $\theta_{\nu}$ 
translates into a significant uncertainty in $L_{\nu}$.

The oscillation sensitivity can be improved by incorporating
information on the $L_{\nu}/E_{\nu}$ resolution into the oscillation analysis.
An example of this technique is the 
Super-Kamiokande L/E analysis~\cite{superk1},
which uses a Monte Carlo simulation to calculate 
the average $L_{\nu}/E_{\nu}$ resolution on a 2D grid 
of reconstructed neutrino energy and zenith angle.
A sample of high resolution events is then selected
by cutting out the regions of energy and angle with 
an average resolution less than 70\%.
The resulting $L_{\nu}/E_{\nu}$ distribution 
of selected events is seen to exhibit a 
characteristic neutrino oscillation dip.

This paper describes a Bayesian technique 
for estimating the $L_{\nu}/E_{\nu}$ resolution
of atmospheric $\nu_{\mu}$ and $\overline{\nu}_{\mu}$ 
neutrinos on an event-by-event basis. 
For each event,
a probability distribution function (PDF) 
in $\log_{10}(L_{\nu}/E_{\nu})$ is calculated 
by combining the measured properties of the 
emitted muon and hadronic system in the event 
with a Monte Carlo simulation of the atmospheric neutrino spectrum,
neutrino interaction kinematics, and detector resolution.
The $L_{\nu}/E_{\nu}$ resolution is then taken as 
the RMS of this PDF. In the oscillation analysis,
all selected atmospheric neutrino events are included,
with events binned according to their resolution.

The Bayesian technique is demonstrated using a simulated
atmospheric neutrino data set from the MINOS experiment.
For the MINOS atmospheric neutrino analysis,
the separation of events into bins of $L_{\nu}/E_{\nu}$ resolution 
is found to yield a significant improvement in the 
oscillation sensitivity.

\section{MINOS Simulation and Reconstruction}
\label{MINOS}

The MINOS Far Detector~\cite{minosdetector} is a 5.4\,kton 
tracking calorimeter located 705\,m underground in the 
Soudan mine, MN, USA. Its large mass and underground location
enable MINOS to measure atmospheric $\nu_{\mu}$ and $\overline{\nu}_{\mu}$
disappearance due to neutrino oscillations.
Atmospheric neutrino $\nu_{\mu}$ and $\overline{\nu}_{\mu}$ CC interactions 
in the MINOS detector are identified by the presence of a muon track 
with a contained-vertex or an upward-going trajectory. 
The detector is also magnetised, allowing muon charge-sign 
to be determined from track curvature. This information is used to 
distinguish between $\nu_{\mu} + N \rightarrow \mu^{-} + X$ 
and $\overline{\nu}_{\mu} + N \rightarrow \mu^{+} + X$ CC interactions.

The analysis presented here uses simulated
contained-vertex atmospheric neutrino interactions
in the MINOS detector.
The MINOS Monte Carlo simulation 
uses the Bartol 3D~\cite{bartol3D} calculation of atmospheric neutrino fluxes
and the {\sc neugen}~\cite{neugen3} model of neutrino cross-sections.
A {\sc geant3}~\cite{geant3} detector simulation tracks the final-state particles
and provides a full description of the detector response and readout.
The simulation is used to generate a sample of 
contained-vertex atmospheric neutrino interactions 
corresponding to a total exposure of 193,000 kton-years.

A series of dedicated algorithms are used to reconstruct 
the muon tracks and hadronic showers on the assumption
that they are produced by atmospheric neutrino interactions~\cite{blake}.
For muon tracks, the propagation direction of the muon 
along the track is first determined using timing information. 
A Kalman Filter algorithm 
is then used to determine the muon trajectory through the detector~\cite{marshall}.
For fully contained muons, which stop in the detector, 
the muon momentum is calculated from the track range; 
for partially contained muons, which exit the detector, 
the momentum is calculated from the track curvature.
For reconstructed showers, the total hadronic energy 
is calculated from the visible energy in the shower.

The atmospheric neutrino energy and direction are calculated 
from the reconstructed muon track and hadronic shower.
The neutrino energy is taken to be the sum of the
muon and shower energy; and the neutrino direction 
is taken to be the muon direction. 
The neutrino propagation distance is calculated by
assuming a fixed 15\,km neutrino production height 
in the atmosphere.

A set of selection requirements are applied that identify 
contained-vertex muon tracks produced by atmospheric 
neutrino interactions~\cite{chapman}. 
This yields a sample of 5.2 million simulated events 
which are used for this analysis.
The $\nu_{\mu}$ and $\overline{\nu}_{\mu}$ CC component 
forms 94\% of the selected sample of events, with the remaining
6\% composed of the NC and $\nu_{e}+\overline{\nu}_{e}$ CC backgrounds,
which do not oscillate in the two-flavour model.
The predicted event rates in the MINOS detector are
27 events per kton-year in the absence of oscillations, 
and 19 events per kton-year using representative oscillation 
parameters of $|\Delta m^{2}| = 2.32 \times10^{-3} \mbox{eV}^{2}$
and $\sin^2 2\theta = 1.0$,
which are assumed throughout this paper.

\begin{figure*}
   \centering
   \includegraphics[width=\textwidth]{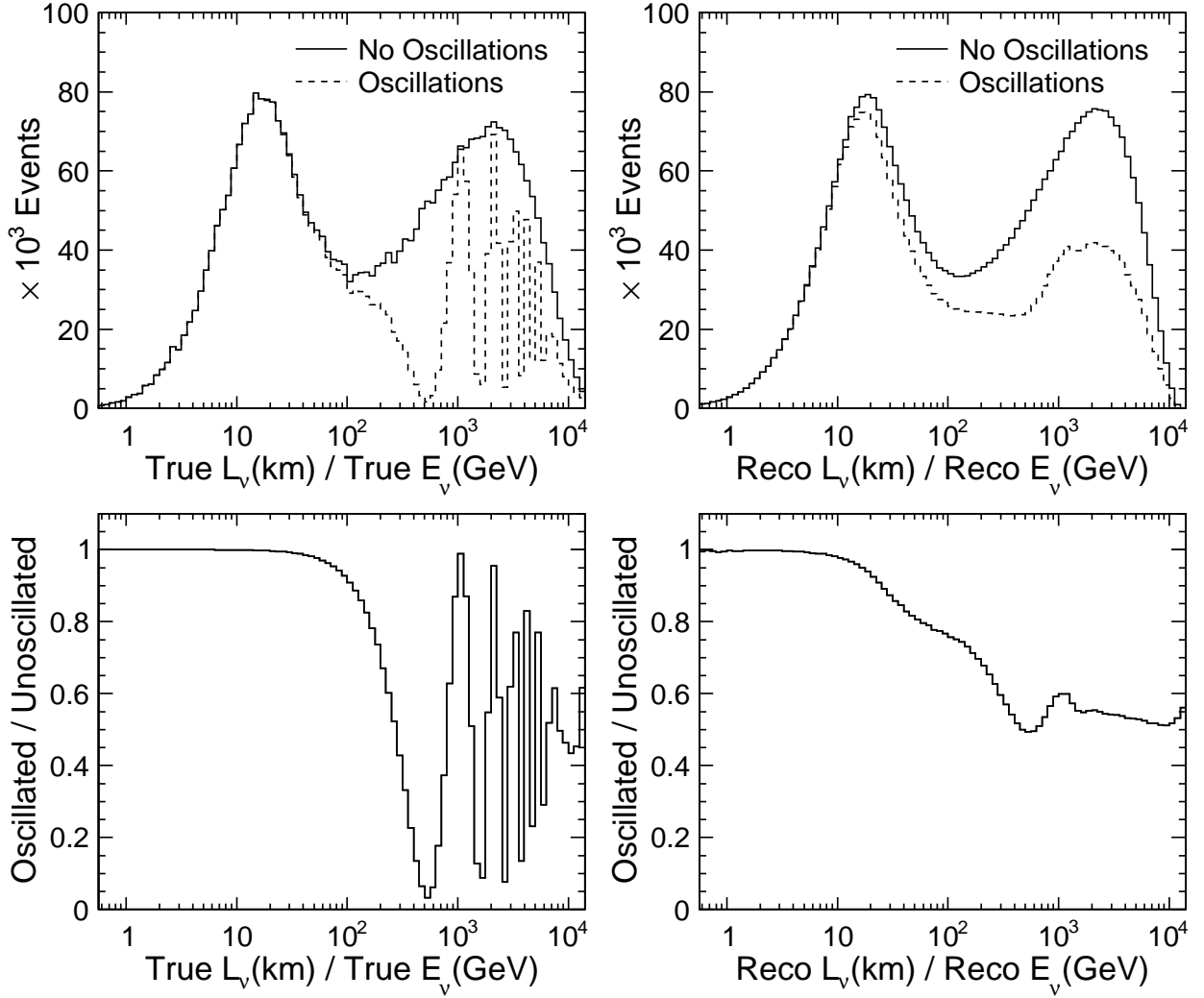}
   \caption{\label{fig_truerecologle}
     The top panels show the true and reconstructed $L_{\nu}/E_{\nu}$ distributions
     for simulated contained-vertex atmospheric neutrinos in the MINOS detector. 
     In each case, the solid line shows the prediction without oscillations
     and the dotted line shows the oscillated prediction,
     using representative oscillation parameters of $\Delta m^{2}=2.32 \times 10^{-3} \mbox{eV}^{2}$ 
     and $\sin^{2}2\theta=1.0$. The bottom panels show the ratios
     of the distributions calculated with and without oscillations.
     The first oscillation maximum is visible in the ratio of
     reconstructed $L_{\nu}/E_{\nu}$ distributions as a dip at $\log_{10}(L_{\nu}/E_{\nu})\approx2.7$.
     However, at higher values of $L_{\nu}/E_{\nu}$, 
     the oscillations are smeared out.
     (Note: the up-turn at very high $L_{\nu}/E_{\nu}$,
     where the lowest energy and hence shortest events occur,
     is due to an increased contamination of downward-going neutrinos 
     mis-identified as upward-going neutrinos.)}
 \end{figure*}

Fig.~\ref{fig_truerecologle} shows the
true and reconstructed $\log_{10}(L_{\nu}/E_{\nu})$ distributions,
plotted with and without oscillations,
for those atmospheric neutrinos
that pass the selection cuts. 
For these plots, and throughout this paper,
the quantities $L_{\nu}$ and $E_{\nu}$ are measured in 
units of km and GeV, respectively.
In the true oscillated distribution,
there is a clear dip at $\log_{10}(L_{\nu}/E_{\nu}) \approx 2.7$,
corresponding to the initial oscillation maximum,
and a clear oscillatory structure at higher values 
of $\log_{10}(L_{\nu}/E_{\nu})$.
In the reconstructed distribution,
the oscillations are smeared by detector resolution,
with only the first oscillation dip clearly visible.

\section{Bayesian L/E Analysis}
\label{BayesianAnalysis}

The aim of the Bayesian analysis is to calculate a PDF in $\log_{10}(L/E)$ 
for each event, based on its observed kinematics.
For MINOS atmospheric neutrinos, these kinematic observables are: 
the reconstructed muon energy, $E^{R}_{\mu}$,
and muon direction, $\hat{p}^{R}_{\mu}$;
and the reconstructed shower energy, $E^{R}_{shw}$.
Each event is also tagged as $q^{R}_{\nu}=(\nu_{\mu}$,$\overline{\nu}_{\mu})$,
as determined by the reconstructed muon charge-sign. 
The corresponding true values for these variables
are denoted $E_{\mu}$, $\hat{p}_{\mu}$, $E_{shw}$ 
and $\mbox{q}_{\nu}$ respectively.

The required PDF, $P$, can be written as follows:

  \begin{equation}
    P = P \left( \hat{p}_{\nu}, E_{\nu}, q_{\nu} \mid \hat{p}_{\mu}^{R}, E_{\mu}^{R}, E_{shw}^{R}, q_{\nu}^{R} \right),
  \label{BayesOne}
  \end{equation}

\noindent
where $\hat{p}_{\nu}$ is the neutrino direction, 
which can be converted into the propagation distance, $L_{\nu}$.

Using Bayes' theorem, with the normalisation $P(\hat{p}_{\mu}^{R},E_{\mu}^{R},E_{shw}^{R},q_{\nu}^{R})=1$, 
Eq.~\ref{BayesOne} can be re-written 
in the following way:

 \begin{equation}
 \begin{split}
   P & = P \left( \hat{p}_{\mu}^{R}, E_{\mu}^{R}, E_{shw}^{R}, q_{\nu}^{R} \mid \hat{p}_{\nu}, E_{\nu}, q_{\nu} \right)\\
     & \times P \left( \hat{p}_{\nu}, E_{\nu}, q_{\nu} \right),
 \end{split}
 \label{BayesTwo}
 \end{equation}

\noindent
where the term $P(\hat{p}_{\nu},E_{\nu},q_{\nu})$ is the Bayesian prior,
and gives the expected distributions of $\hat{p}_{\nu}$ and $E_{\nu}$
for neutrinos and antineutrinos.

Eq.~\ref{BayesTwo} is further modified by introducing PDFs 
relating the true and reconstructed event kinematics.
The result is as follows:
 
 \begin{equation}
 \begin{split}
   P & = P \left( \hat{p}_{\mu}^{R}, E_{\mu}^{R}, E_{shw}^{R}, q_{\nu}^{R} \mid \hat{p}_{\mu}, E_{\mu}, E_{shw}, q_{\nu} \right)\\ 
     & \times P \left( \hat{p}_{\mu}, E_{\mu},E_{shw} \mid \hat{p}_{\nu}, E_{\nu}, q_{\nu} \right)\\ 
     & \times P \left( \hat{p}_{\nu}, E_{\nu}, q_{\nu} \right).
 \end{split} 
 \label{BayesThree}
 \end{equation}

In Eq.~\ref{BayesThree}, the first term, which is written as
$P(\hat{p}_{\mu}^{R},E_{\mu}^{R},E_{shw}^{R},q_{\nu}^{R}|\hat{p}_{\mu},E_{\mu},E_{shw},q_{\nu})$,
contains a set of resolution functions connecting the reconstructed
muon and shower kinematics to the underlying true muon and shower distributions.
This term also contains the relative selection 
efficiency for atmospheric neutrinos as a function of energy.
The second term,
$P(\hat{p}_{\mu},E_{\mu},E_{shw}|\hat{p}_{\nu},E_{\nu},q_{\nu})$,
contains a set of kinematic distributions connecting the 
true muon and shower kinematics to the underlying neutrino interaction kinematics.
This term incorporates the relative cross-sections and kinematic distributions
for quasi-elastic (QE), resonance (RES), and deep-inelastic (DIS) 
$\nu_{\mu}$ and $\overline{\nu}_{\mu}$ CC interactions.
The third term, $P(\hat{p}_{\nu},E_{\nu},q_{\nu})$, is the Bayesian prior, 
described above.
Taken together, the three terms combine the relative probability
of an atmospheric neutrino interaction at a given energy and angle,
with the distribution of kinematic observables produced by 
the neutrino interaction.

The following subsections describe the calculation of each
of the terms in Eq.~\ref{BayesThree}, 
along with the approximations made in their calculations.

\subsection{Resolution Functions}

The resolution functions translate the underlying distributions 
of true muon and shower variables into the distributions 
of reconstructed variables.
To calculate the resolution functions, it is assumed that
each observable is measured independently of the others.
In this approximation, the resolution functions decouple
into single-variable parameterisations.
For the muon and shower energy, these are given by
$P(E_{\mu}^{R}|E_{\mu})$ and $P(E_{shw}^{R}|E_{shw},q_{\nu})$ respectively,
where separate shower resolution functions are calculated 
for neutrinos and antineutrinos. 
For the muon direction, a perfect resolution is assumed.
This is a reasonable approximation, since the average
angular resolution of $1-2$~degrees for reconstructed muon tracks 
is significantly smaller than the average angle between the neutrino
and emitted muon. Therefore, the muon direction is
fixed by setting $P(\hat{p}_{\mu}^{R}|\hat{p}_{\mu})=\delta(\hat{p}_{\mu}^{R}-\hat{p}_{\mu})$.
For the muon charge-sign, the set of probabilities $P(q_{\nu}^{R}|q_{\nu})$
are approximated using a similar method. 
The atmospheric neutrino selection criteria are used 
to identify muons that have a well-measured curvature. 
For these events, a perfect charge-sign reconstruction
is assumed, given by $P(q_{\nu}^{R}|q_{\nu})=\delta(q_{\nu}^{R}-q_{\nu})$.
The vast majority of the contained-vertex events 
fall into this category. For those remaining events,
it is assumed that $P(q_{\nu}^{R}|q_{\nu})=1/2$.

With these approximations, the overall resolution
function can be written as follows:

 \begin{equation}
 \begin{split}
   & P \left( \hat{p}_{\mu}^{R}, E_{\mu}^{R}, E_{shw}^{R}, q_{\nu}^{R} \mid \hat{p}_{\mu}, E_{\mu}, E_{shw}, q_{\nu} \right)\\ 
   & = \delta \left( \hat{p}_{\mu}^{R} - \hat{p}_{\mu} \right) \times P \left( q_{\nu}^{R} \mid q_{\nu} \right)\\ 
   & \times P \left( E_{\mu}^{R} \mid E_{\mu} \right) \times P \left( E_{shw}^{R} \mid E_{shw}, q_{\nu} \right)\\
 \end{split}
 \end{equation}

To calculate $P(E_{\mu}^{R}|E_{\mu})$ and $P(E_{shw}^{R}|E_{shw},q_{\nu})$, 
the 5.2 million Monte Carlo events are used to populate 2D distributions
in the reconstructed and true energy of muon tracks and hadronic showers. 
The distributions are then parameterised to create resolution functions 
that return a relative probability for each combination of reconstructed 
and true energy.
For muon tracks, separate parameterisations are constructed
for stopping and exiting muons, which correspond to 
fully contained and partially contained events, respectively. 
For hadronic showers, separate parameterisations are constructed 
for $\nu_{\mu}$ CC QE, $\nu_{\mu}$ CC RES+DIS, 
$\overline{\nu}_{\mu}$ CC QE and $\overline{\nu}_{\mu}$ CC RES+DIS 
neutrino interactions.

\begin{figure}
   \centering
   \includegraphics[width=\columnwidth]{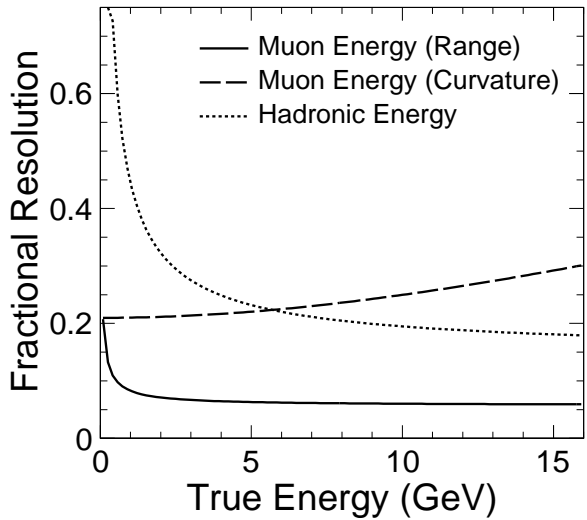}
   \caption{\label{fig_resolution}
    Parameterised energy resolution functions for reconstructed 
    atmospheric neutrino $\nu_{\mu}$ and $\overline{\nu}_{\mu}$ CC 
    interactions in the MINOS detector. 
    The fractional energy resolutions are plotted as a function of the true energy for: 
    the muon energy calculated from range for fully contained muons (solid line);
    the muon energy calculated from curvature for partially contained muons (dashed line);
    and the hadronic shower energy (dotted line).}
 \end{figure}

Fig.~\ref{fig_resolution} shows a parameterisation 
of the average fractional resolutions as a function of true energy,
for the muon energy, determined from either range or curvature,
and for the hadronic shower energy. 
The resolutions are averaged over all the selected atmospheric neutrino
$\nu_{\mu}$ and $\overline{\nu}_{\mu}$ CC interactions.
For fully contained events, the muon energy 
is determined from track range with 
a typical resolution of a few percent.
However, the muon energy from track curvature
and the hadronic shower energy both have
significantly worse resolutions.
Therefore, the overall neutrino energy resolution is better
for fully contained than partially contained events,
and also better for neutrino interactions with smaller inelasticities.

 \begin{figure*}
   \centering
   \includegraphics[width=\textwidth]{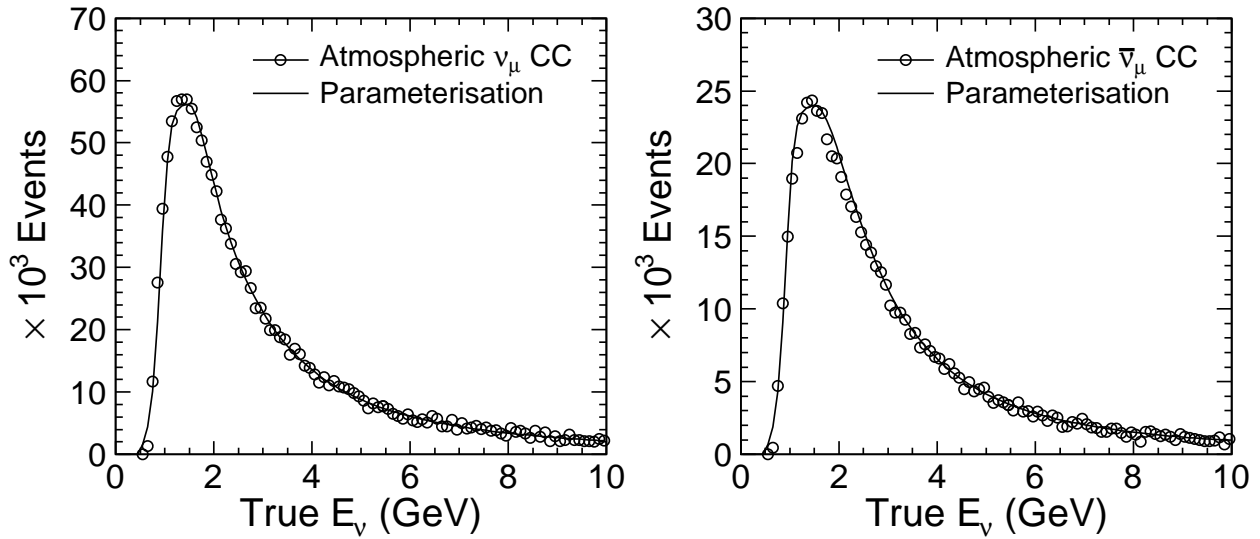}
   \caption{\label{fig_bayesprior}
     Distributions of true neutrino energy for Monte Carlo
     atmospheric $\nu_{\mu}$ (left) and $\overline{\nu}_{\mu}$ (right) 
     CC interactions that satisfy the event selection cuts. 
     In each plot, the open circles show the distributions of 
     selected Monte Carlo events, and the solid line shows the 
     parameterisations used as priors in the Bayesian L/E analysis.
     At low energies, the spectrum rises sharply as the selection 
     efficiency increases rapidly from zero; at higher energies,
     the spectrum falls away, reflecting the underlying 
     atmospheric neutrino energy spectrum.}
 \end{figure*}

\subsection{Kinematic Distributions}

A set of kinematic distributions are used to calculate 
PDFs of the true muon and shower observables
for a given neutrino energy and direction.
To construct the kinematic distributions,
the {\sc neugen} simulation is used to generate a sample of 
1 billion $\nu_{\mu}$ and $\overline{\nu}_{\mu}$ CC interactions
with a uniform energy spectrum.
The simulated interactions are separated into neutrinos and antineutrinos
and are binned according to their interaction type (QE, RES, DIS).
In each bin, a 3D PDF in neutrino energy, $E_{\nu}$, 
and the kinematic variables, $W^{2}$ and $y$, is populated,
where $W^{2}$ is the invariant mass squared of the final-state hadronic system,
and $y$ is the inelasticity of the interaction.
By neglecting the effects of Fermi momentum in the interactions, 
each combination of kinematic variables ($E_{\nu}$, $W^{2}$, $y$)
can be mapped on to a single muon energy, hadronic energy, 
and relative angle between the neutrino and muon.
Therefore, for a given neutrino energy and direction, 
PDFs of the true muons and shower observables
($\hat{p}_{\mu}, E_{\mu}, E_{shw}$) 
can be calculated by marginalising the kinematic distributions over $W^{2}$ and $y$,
weighting each bin of interaction type
according to its relative probability as a
function of neutrino energy.

\subsection{Bayesian Prior}

The Bayesian prior gives the expected distributions 
in energy and angle for $\nu_{\mu}$ and $\overline{\nu}_{\mu}$ CC interactions.
For the analysis presented here, the distributions
are obtained from the Monte Carlo simulation.
The expected energy distribution, $P(E_{\nu},q_{\nu})$,
is found by calculating the product of the flux and total cross-section
for neutrinos and antineutrinos. 
For the angular distribution, it is assumed that the
incident flux of atmospheric neutrinos is isotropic. 
This is a reasonable approximation, since the selected 
events have an average energy of $>1$\,GeV. 
In this energy region, the atmospheric 
$\nu_{\mu}$ and $\overline{\nu}_{\mu}$ flux
is approximately uniform as a function of
the neutrino zenith angle~\cite{bartol3D}.

The energy distributions of 
$\nu_{\mu}$ and $\overline{\nu}_{\mu}$ CC interactions
is combined with the selection efficiency, 
$\epsilon^{R}(E_{\nu},q_{\nu})$, 
which has a strong energy dependence.
Fig.~\ref{fig_bayesprior} shows the resulting
energy distributions, $P^{R}(E_{\nu},q_{\nu})=P(E_{\nu},q_{\nu})\times\epsilon^{R}(E_{\nu},q_{\nu})$, 
for selected neutrinos and antineutrinos.
These distributions are parameterised for
the Bayesian $L_{\nu}/E_{\nu}$ analysis. In each case, 
the distributions rise sharply at low energies, 
as the selection efficiency increases rapidly from zero;
the distributions then fall away again at higher energies,
reflecting the underlying atmospheric neutrino energy spectrum.

The calculated $L_{\nu}/E_{\nu}$ resolutions are found 
to be insensitive to the exact choice of Bayesian prior,
particularly the best $L_{\nu}/E_{\nu}$ resolutions,
where only a small region in neutrino energy and zenith angle
contributes significantly to the Bayesian PDF.
When a flat energy spectrum is used as a prior,
similar results are obtained.

 \begin{figure*}
   \centering
   \includegraphics[width=0.49\textwidth]{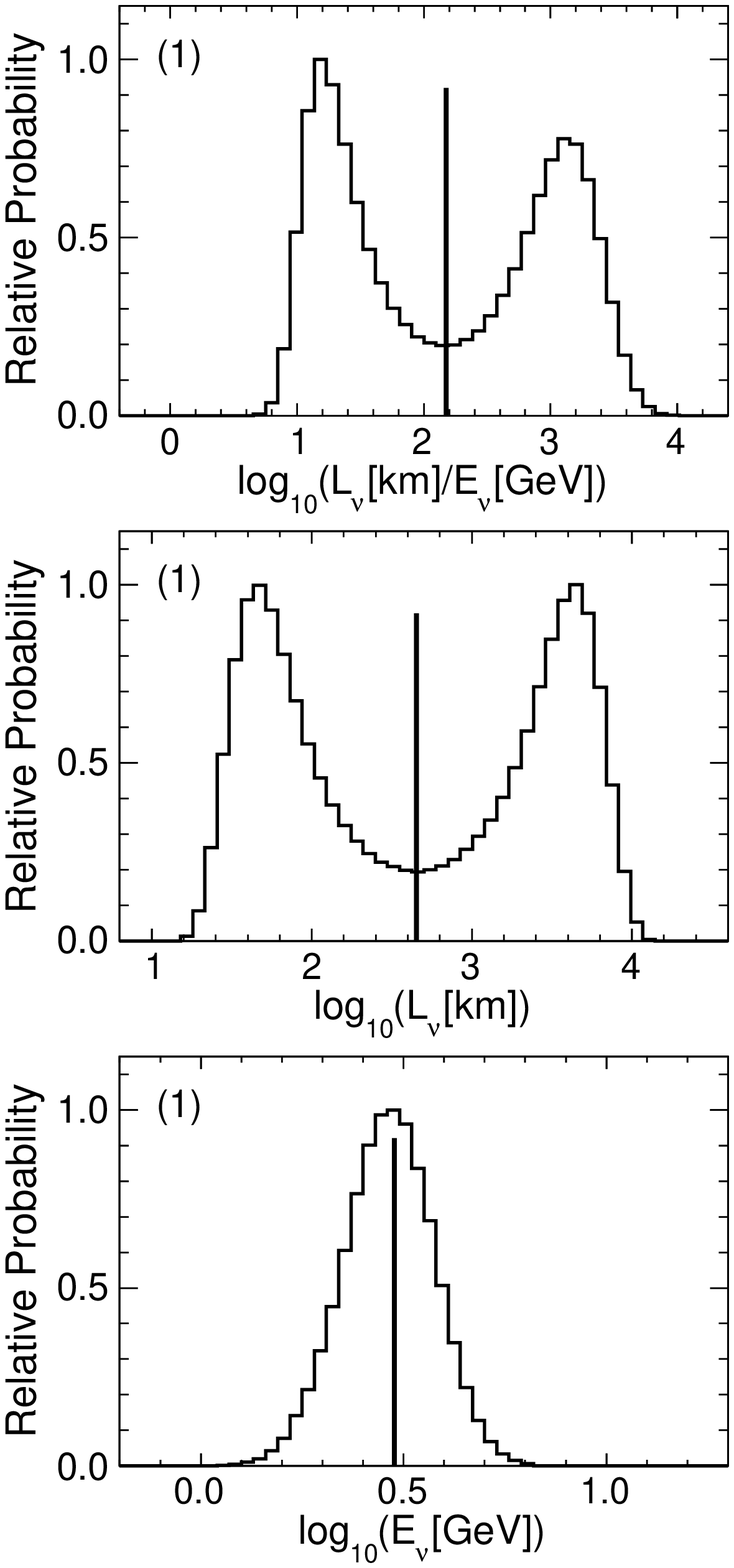}
   \includegraphics[width=0.49\textwidth]{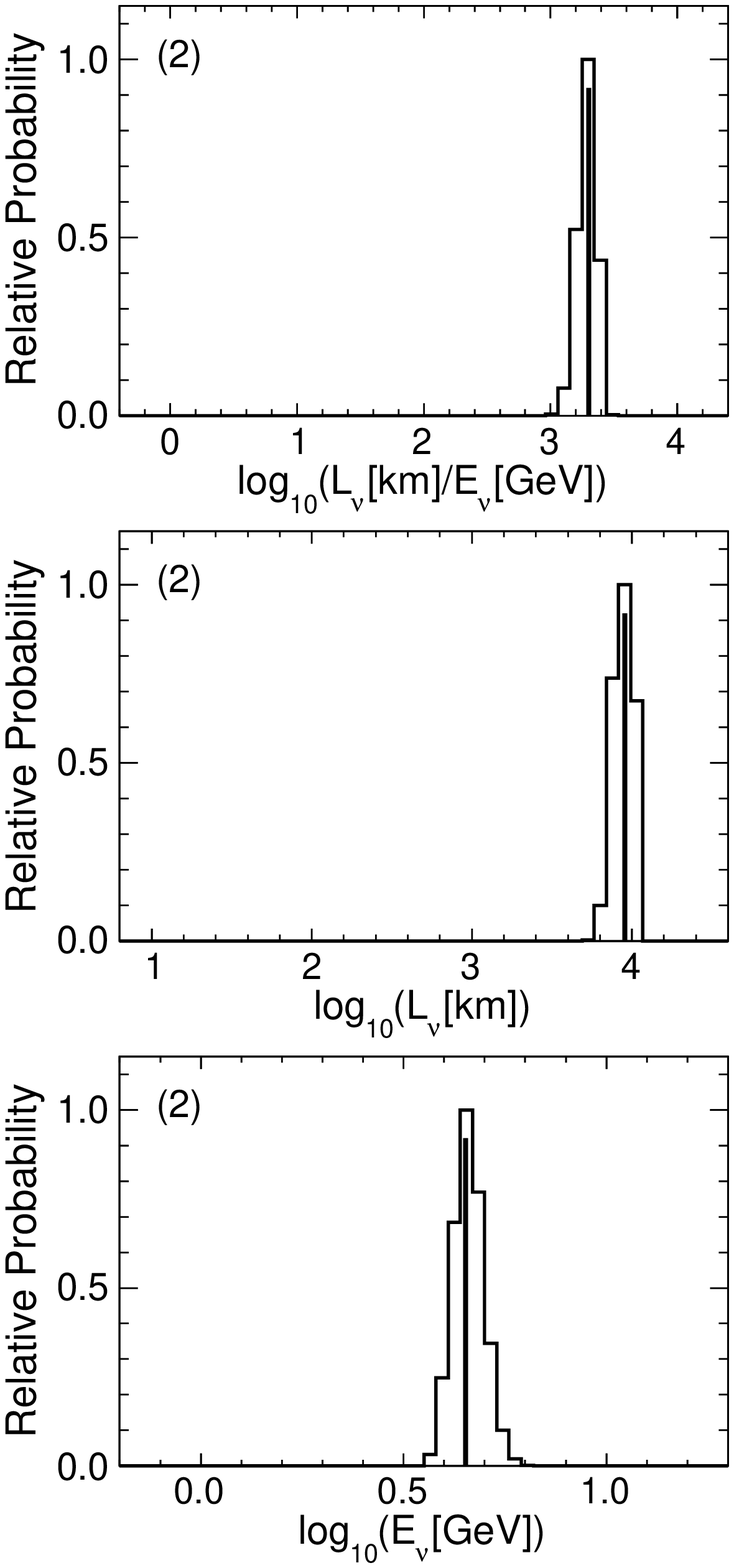}
   \caption{\label{fig_bayesexamples}
    Examples of Bayesian PDFs in $\log_{10}(L_{\nu}/E_{\nu})$, $\log_{10}(L_{\nu})$ and $\log_{10}(E_{\nu})$
    for two reconstructed atmospheric neutrinos. 
    The left panels show a partially contained event with 
    reconstructed energies of $E_{\mu}^{R}=2$\,GeV and $E_{shw}^{R}=1$\,GeV.
    The muon direction is perfectly horizontal and the charge-sign is negative.
    The $L_{\nu}/E_{\nu}$ resolution is calculated to be $\sigma_{\log(L_{\nu}/E_{\nu})}=0.87$.
    The right panels show a fully contained event with 
    reconstructed energies of $E_{\mu}^{R}=4$\,GeV and $E_{shw}^{R}=0.5$\,GeV,
    The muon direction is directed $45^{\circ}$ upwards, and the charge-sign is positive.
    The $L_{\nu}/E_{\nu}$ resolution is calculated to be $\sigma_{\log(L_{\nu}/E_{\nu})}=0.07$.
    In each plot, the vertical lines indicate the reconstructed values,
    and the histograms give the calculated PDFs.}
 \end{figure*}

\subsection{Calculating the Bayesian PDF}

For each event, a posterior PDF in $\log_{10}(L/E)$
is calculated by multiplying the resolution functions, 
kinematic distributions and Bayesian prior,
and marginalising over the kinematic variables $E_{\nu}$, $W^{2}$, $y$.
At each point in this parameter space,
the true muon momentum and hadronic energy is calculated,
along with the relative angle between the muon and incident neutrino.
The true muon direction is fixed on the reconstructed muon direction, 
and an additional integral is then performed over an angle, $\phi$, 
which rotates the neutrino around this direction. 
At each point in the integral, the true values of $L_{\nu}$ and $E_{\nu}$,
and therefore $\log_{10}(L_{\nu}/E_{\nu})$, are calculated.
A weight is assigned as given by the product of the input PDFs, 
and the value of $\log_{10}(L_{\nu}/E_{\nu})$ 
is entered into the posterior PDF with this weight.
The values of $\log_{10}(L_{\nu})$ and $\log_{10}(E_{\nu})$ 
are entered into a separate set of PDFs,
providing information on the relative resolution 
of $L_{\nu}$ and $E_{\nu}$ for each event.
After the full PDF in $\log_{10}(L_{\nu}/E_{\nu})$ has been calculated,
its RMS value, labelled $\sigma_{\log(L_{\nu}/E_{\nu})}$, 
is taken as the $L_{\nu}/E_{\nu}$ resolution of the event.

\subsection{Examples}

Fig.~\ref{fig_bayesexamples} shows examples 
of PDFs in $\log_{10}(L_{\nu})$, $\log_{10}(E_{\nu})$ and $\log_{10}(L_{\nu}/E_{\nu})$ 
for two reconstructed atmospheric neutrinos.
The first event is a partially contained muon 
with reconstructed energies of $E^{R}_{\mu}=2$\,GeV 
and $E^{R}_{shw}=1$\,GeV. 
The reconstructed muon track direction, $\hat{p}^{R}_{\mu}$,
is perfectly horizontal, and the reconstructed 
muon charge-sign is negative, $q^{R}=-1$,
implying a $\nu_{\mu}$ CC interaction.
For this event, the resulting PDF in $\log_{10}(L_{\nu}/E_{\nu})$
has a double-peaked structure, arising from
the rapid variation in the propagation distance, $L_{\nu}$,
with the neutrino zenith angle close to the horizon.
The resolutions on $L_{\nu}$ and $E_{\nu}$ 
are calculated to be $\sigma_{\log(L_{\nu})}=0.86$ 
and $\sigma_{\log(E_{\nu})}=0.11$, respectively,
and the overall resolution is $\sigma_{\log(L_{\nu}/E_{\nu})}=0.87$.
This is an example of a low resolution event,
where the broad $\log_{10}(L_{\nu})$ distribution
makes the dominant contribution to the overall
$L_{\nu}/E_{\nu}$ resolution.

The second event is a fully contained muon that has reconstructed
energies of $E^{R}_{\mu}=4$\,GeV and $E^{R}_{shw}=0.5$\,GeV.
The reconstructed muon direction is directed upwards at an 
angle of $45^{\circ}$ to the horizontal,
and the reconstructed muon charge-sign is positive, $q^{R}=+1$,
implying a $\overline{\nu}_{\mu}$ CC interaction.
Since the muon direction is significantly above the horizon, 
the variations in $L_{\nu}$ as a function of zenith angle are small,
giving a narrow PDF in $\log_{10}(L_{\nu})$, with $\sigma_{\log(L_{\nu})}=0.06$.
Since the event is fully contained, 
the neutrino energy is also well-measured and the PDF in $\log_{10}(E_{\nu})$ 
returns a small resolution of $\sigma_{\log(E_{\nu})}=0.04$.
The two distributions combine to give a narrow PDF in $\log_{10}(L_{\nu}/E_{\nu})$,
with an overall $L_{\nu}/E_{\nu}$ resolution of $\sigma_{\log(L_{\nu}/E_{\nu})}=0.07$.
This an example of a high resolution event.

\section{Separation of Events by L/E Resolution}
\label{ResolutionBinning}

\begin{figure}
   \centering
   \includegraphics[width=\columnwidth]{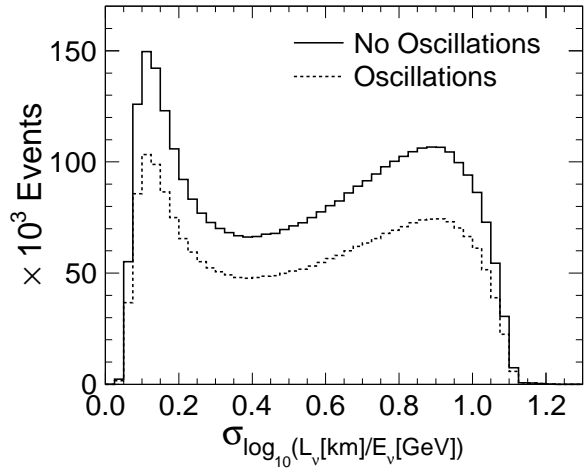}
   \caption{\label{fig_bayesresolution}
     Distribution of Bayesian L/E resolution, $\sigma_{\log(L/E)}$,
     for the simulated atmospheric neutrino sample.
     The solid histogram shows the predicted distribution in the absence
     of oscillations; the dotted line shows the predicted distribution
     for representative oscillation parameters of 
     $\Delta m^{2} = 2.32 \times 10^{-3} \mbox{eV}^{2}$
     and $\sin^{2} 2\theta = 1.0$.}
 \end{figure}

Fig.~\ref{fig_bayesresolution} shows the distribution of 
$\sigma_{\log(L_{\nu}/E_{\nu})}$ for the simulated event sample,
plotted with and without oscillations.
There is a substantial spread of $\sigma_{\log(L_{\nu}/E_{\nu})}$ values
across the event sample, 
corresponding to $\sim 25$\% of the overall spread of 
the reconstructed neutrino $\log_{10}(L_{\nu}/E_{\nu})$ distribution.
The lowest values of  $\sigma_{\log(L_{\nu}/E_{\nu})}$ are roughly
an order of magnitude smaller than the highest values.
The shape of the predicted $\sigma_{\log(L_{\nu}/E_{\nu})}$ distribution 
is also approximately independent of oscillations.

  \begin{table*}
  \begin{center}
  \begin{tabular}{cccc}
  \toprule
       Bin number & Two resolution bins & Four resolution bins & Eight resolution bins \\
  \midrule
      1 & $0.00\leq\sigma_{\log(L/E)}<0.50$ & $0.00\leq\sigma_{\log(L/E)}<0.25$ & $0.000\leq\sigma_{\log(L/E)}<0.125$ \\
      2 & $0.50\leq\sigma_{\log(L/E)}<1.50$ & $0.25\leq\sigma_{\log(L/E)}<0.50$ & $0.125\leq\sigma_{\log(L/E)}<0.250$ \\
      3 &              $-$                 & $0.50\leq\sigma_{\log(L/E)}<0.75$ & $0.250\leq\sigma_{\log(L/E)}<0.375$ \\
      4 &              $-$                 & $0.75\leq\sigma_{\log(L/E)}<1.50$ & $0.375\leq\sigma_{\log(L/E)}<0.500$ \\
      5 &              $-$                 &              $-$                 & $0.500\leq\sigma_{\log(L/E)}<0.625$ \\
      6 &              $-$                 &              $-$                 & $0.625\leq\sigma_{\log(L/E)}<0.750$ \\
      7 &              $-$                 &              $-$                 & $0.750\leq\sigma_{\log(L/E)}<0.950$ \\
      8 &              $-$                 &              $-$                 & $0.950\leq\sigma_{\log(L/E)}<1.500$ \\
  \bottomrule
  \end{tabular} 
  \end{center}
  \caption{\label{tab_resolution binning}
     Binning schemes used to separate selected events according to the calculated $L_{\nu}/E_{\nu}$ resolution.}
  \end{table*}

\begin{figure*}
   \centering
   \includegraphics[width=\textwidth]{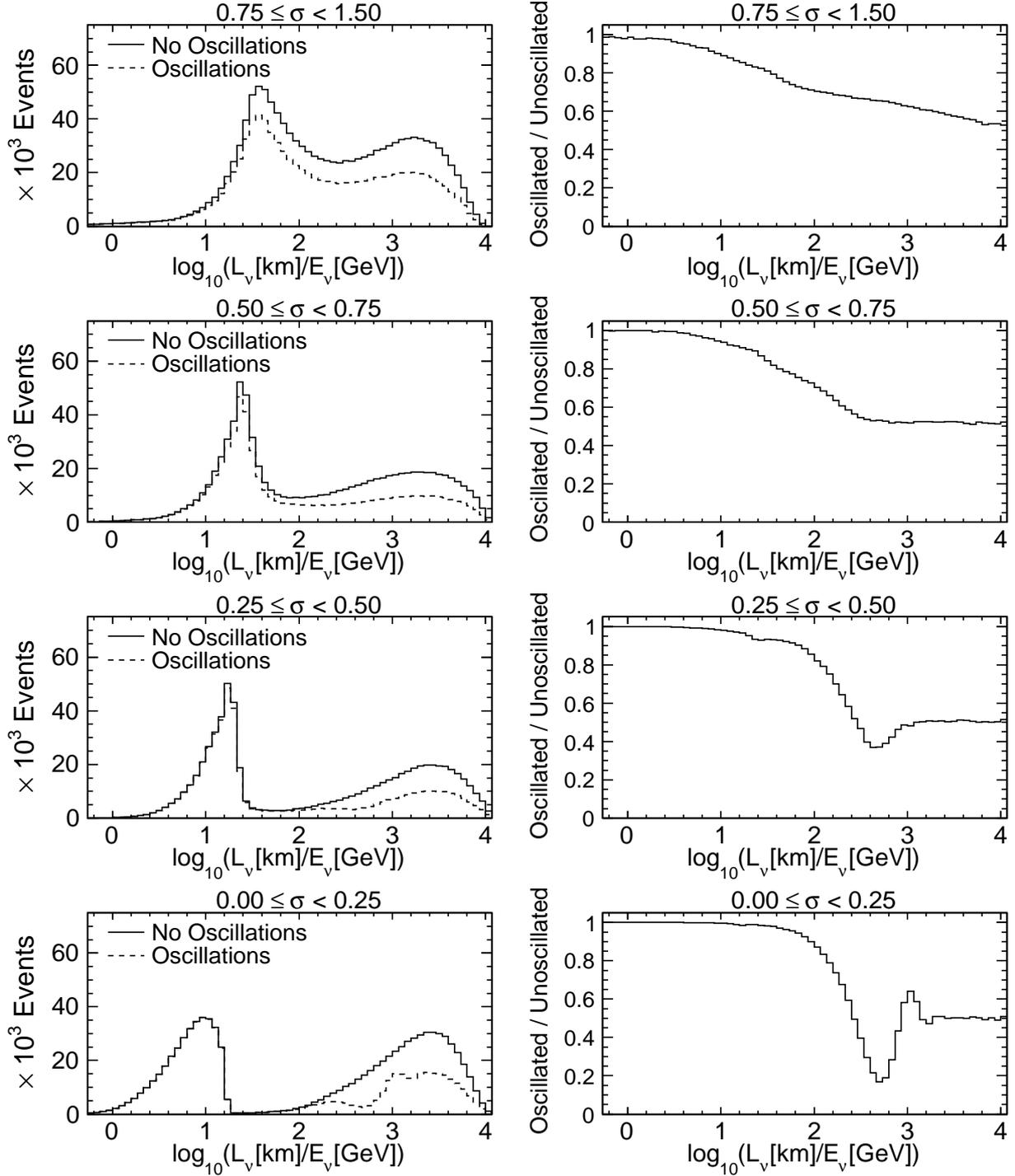}
   \caption{\label{fig_bayeslogle}
     The left panels show the reconstructed $\log(L/E)$ distributions
     for simulated atmospheric neutrinos, 
     plotted with and without oscillations, 
     and separated into the following four bins of resolution:
     $0$$<$$\sigma_{\log(L/E)}$$<$$0.25$; $0.25$$<$$\sigma_{\log(L/E)}$$<$$0.5$;
     $0.5$$<$$\sigma_{\log(L/E)}$$<$$0.75$; $0.75$$<$$\sigma_{\log(L/E)}$$<$$1.5$.
     In each panel, the solid line indicates the prediction in the
     absence of oscillations, and the dotted line gives the
     prediction for representative oscillation parameters of
     $\Delta m^{2} = 2.32 \times 10^{-3} \mbox{ eV}^{2}$
     and $\sin^{2} 2\theta = 1.0$.
     The right panels show the ratios of the Monte Carlo
     predictions with and without oscillations.
     The oscillation structure becomes increasingly clear for
     bins of higher resolution, and is most sharply defined 
     in the bin with the highest resolution.}
 \end{figure*}

To calculate oscillation sensitivities, 
the events are separated into bins of $L_{\nu}/E_{\nu}$ resolution,
using the two-bin, four-bin and eight-bin schemes
given in Table~\ref{tab_resolution binning}.
Fig.~\ref{fig_bayeslogle} shows the reconstructed distributions
of $\log_{10}(L_{\nu}/E_{\nu})$ for the four-bin scheme,
calculated with and without neutrino oscillations. 
Also shown are the ratios 
of the distributions with and without oscillations
for each resolution bin. 
The oscillation dip is seen to become increasingly pronounced
with improving resolution,
and is most sharply defined 
in the bin with best resolution. Here,
the ratio initially falls with $\log_{10}(L_{\nu}/E_{\nu})$,
reaching a first minimum at the point of maximum oscillation probability.
As $\log_{10}(L_{\nu}/E_{\nu})$ increases, 
a second oscillation dip is clearly visible,
before the ratio tends to an average value of $1-\frac{1}{2}\sin^{2}2\theta=0.5$,
as the frequency of the oscillations becomes rapid.
Since the highest resolution bin contains a sample of events 
with a better resolved oscillation structure, 
the separation of events into bins of resolution
is expected to yield a significant improvement in 
the oscillation sensitivity.

The shape of the $\log_{10}(L_{\nu}/E_{\nu})$ distribution
in each $L_{\nu}/E_{\nu}$ resolution bin reflects the underlying 
distribution of neutrino energies and angles. 
The events with worse resolution 
are associated with lower energies and angles closer to the horizon,
whereas the events with higher resolution are associated with
higher energies and steeper angles.
Therefore, the central region of the
$\log_{10}(L_{\nu}/E_{\nu})$ distribution,
associated with horizontal neutrinos,
is increasingly suppressed in higher resolution bins.
The oscillation structure is most sharply resolved in
multi-GeV upward-going events, which typically have
the best $L_{\nu}/E_{\nu}$ resolution.

\section{Neutrino Oscillation Sensitivity Study}
\label{OscillationSensitivity}

The impact of $L_{\nu}/E_{\nu}$ resolution binning 
on the oscillation measurement is evaluated 
by performing a maximum likelihood analysis on the reconstructed
$\log_{10}(L_{\nu}/E_{\nu})$ distributions.
The projected oscillation sensitivity is calculated 
first without resolution binning, and then with 
resolution binning using the two-bin, four-bin 
and eight-bin schemes.
 
For this sensitivity study, the Monte Carlo distributions 
are scaled to an equivalent exposure of 37.9 kton-years, 
matching the analysis described in~\cite{minosatmos}.
The selected events are binned according to their reconstructed
muon charge-sign, $q$=($\nu$,$\overline{\nu}$,$X$),
corresponding to neutrinos ($\nu$), antineutrinos ($\overline{\nu}$)
and those events with an ambiguous charge-sign ($X$). 
Each bin of charge-sign is then
separated into the required number of $L_{\nu}/E_{\nu}$ resolution bins.

The oscillation sensitivities are calculated on a 
2D grid in the oscillation parameters $(\Delta m^{2}, \sin^{2} 2\theta)$. 
At each grid point, the reconstructed $\log_{10}(L_{\nu}/E_{\nu})$ distributions
are calculated using the selected Monte Carlo events. 
In addition, a set of predicted distributions are calculated
using representative oscillation parameters of
$\Delta m^{2} = 2.32 \times 10^{-3} \mbox{ eV}^{2}$
and $\sin^{2} 2\theta = 1.0$. 
These distributions are treated as simulated data, 
and used to evaluate the following negative 
log-likelihood function:

\begin{equation}
 \begin{split}
    -\ln {\cal{L}} &= \sum_{q} \mu - n \ln \mu\\
                   &- \sum_{q} \sum_{i,k} n_{ik} \ln \left(f_{ik}\right)\\
                   &+ \sum_{j} \frac{\alpha_j^2}{2\sigma_{\alpha_j}^2}.
  \end{split}
\end{equation}

\noindent
This log-likelihood function is divided into the following terms:

\begin{enumerate}
 \item \textit{Normalisation:}
   The sums $\sum_{q} \mu - n \ln \mu$ represent the
   Poisson probability for observing $n$ total events,
   with a total prediction of $\mu$ events. 
   The sum is taken over the charge-sign 
   bins $q$=($\nu$,$\overline{\nu}$,$X$).
 \item \textit{Shape Term:}
   The terms $\sum_{q} \sum_{i,k} n_{ik} \ln \left(f_{ik} \right)$
   represent the likelihood functions for each of 
   the $\log_{10}(L/E)$ distributions used in the fit. 
   The $i$-sum is taken over the resolution bins; 
   the $k$-sum is taken over the $\log_{10}(L/E)$ bins.
   Within the sum, $n_{ik}$ is the observed number of events
   and $f_{ik}$ is the relative probability in
   the $i^{th}$ and $k^{th}$ bins.
 \item \textit{Systematic Error Term:}
   Systematic effects are incorporated as nuisance parameters, 
   where the shift $\alpha_{j}$ represents the deviation of the $j^{th}$
   systematic parameter from its nominal value.
   A set of penalty terms, $\alpha_j^2/2\sigma_{\alpha_j}^2$,
   are added, where the error $\sigma_{\alpha_j}$
   represents the estimated uncertainty in
   the $j^{th}$ systematic parameter.
   A~total of 10 systematic effects
   are included in the log-likelihood function,
   to account for systematic uncertainties in
   the atmospheric neutrino flux and cross-section calculations~\cite{minosatmos}.
   The log-likelihood function is minimised with 
   respect to each of the systematic parameters.
\end{enumerate}

By evaluating the log-likelihood function at each grid point,
a likelihood surface is constructed in
$(\Delta m^{2}, \sin^{2} 2\theta)$ parameter space. 
The confidence levels (C.L.) on the oscillation parameters
are then calculated assuming Gaussian statistics,
where the two-parameter 68\% C.L. and 90\% C.L. are given 
by the locus of points with log-likelihood values of 
$-\Delta \ln \cal{L} =$ (1.15, 2.30) relative
to the central value at the input oscillation parameters.

\begin{figure}
   \centering
   \includegraphics[width=\columnwidth]{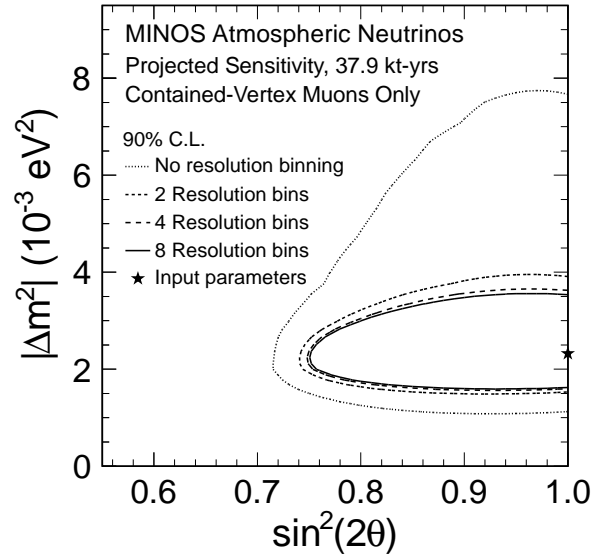}
   \caption{\label{fig_bayessensitivity}
     The projected 90\% confidence limits on the oscillation parameters
     $\Delta m^{2}$ and $\sin^{2} 2\theta$, calculated without 
     resolution binning, and for the cases of two, four and eight bins of resolution.
     The sensitivities are calculated by scaling the 
     contained-vertex atmospheric neutrino Monte Carlo sample 
     to a total exposure of 37.9 kton-years.
     The input
     oscillation parameters are $\Delta m^{2}=2.32\times 10^{-3}\mbox{ eV}^{2}$ 
     and $\sin^{2} 2\theta=1.0$, as indicated by the star.}
 \end{figure}

Fig.~\ref{fig_bayessensitivity} shows the resulting 90\% C.L. contours,
calculated without any resolution binning, 
and for the case of two, four and eight bins of resolution.
The use of resolution binning is found to yield significant improvements
in oscillation sensitivity, particularly for the $\Delta m^{2}$ parameter. 
The sensitivity improves with each doubling in the number of resolution bins, 
with the improvements becoming smaller each time.
A set of single-parameter 90\% C.L. are calculated 
for the each of the oscillation parameters and are found to improve from 
$1.4<|\Delta m^{2}|/10^{-3} \mbox{eV}^{2}<4.9$ and $\sin^{2} 2\theta>0.79$
without using resolution binning,
to $1.7<|\Delta m^{2}|/10^{-3} \mbox{eV}^{2}<3.2$ and $\sin^{2} 2\theta>0.81$
for the case of eight bins of resolution.

\section{Summary}

For atmospheric neutrino $\nu_{\mu}$ and $\overline{\nu}_{\mu}$ CC interactions,
the $L_{\nu}/E_{\nu}$ resolution is crucially
important in determining the sensitivity to oscillations, 
but varies significantly from event to event. 
This paper has described a Bayesian technique for 
estimating the $L_{\nu}/E_{\nu}$ resolution on an event-by-event basis,
which enables an event sample to be separated into bins of resolution.
The technique has been demonstrated 
using simulated atmospheric neutrino data from the MINOS experiment. 
The selected events
are binned as a function of $\log_{10}(L_{\nu}/E_{\nu})$ 
and also by $L_{\nu}/E_{\nu}$ resolution.
The resolution binning takes full advantage of high resolution events,
which sharply resolve the oscillations.
It also allows low resolution events to be included in the analysis,
which contribute to the oscillation sensitivity by providing
information on the relative rate of upward-going 
and down-going neutrinos.
By separating events into bins of $L_{\nu}/E_{\nu}$ resolution,
a significant improvement in the oscillation sensitivity
can be achieved.

\section*{Acknowledgements}

The authors gratefully acknowledge the support
of the MINOS collaboration and would like to thank 
Jon Urheim, Brett Viren and Anna Holin
for their helpful suggestions and comments on this paper.
This work was supported by the STFC, UK.

\bibliographystyle{elsarticle-num}

\bibliography{bayes_paper}

\end{document}